# 'Super-apps' and the Digital Markets Act

Simonetta Vezzoso, Trento University

**Abstract**

The Digital Markets Act (DMA) aims to ensure contestability and fairness in digital markets, particularly focusing on regulating Big Tech companies. The paper explores the DMA's capacity to address both current and future challenges in digital market contestability and fairness, spotlighting the trend towards platform integration and the potential rise of "super-apps" akin to WeChat and KakaoTalk. Specifically, it investigates WhatsApp, owned by Meta, as a gatekeeper that might expand its service offerings, integrating additional functionalities like AI and metaverse technologies. The paper discusses whether the DMA's obligations, such as mandated interoperability and data portability, can mitigate the emergent risks to market fairness and contestability from such integrations. Despite recognizing that the DMA has the potential to address many issues arising from platform integration, it suggests the necessity for adaptability and a complementary relationship with traditional antitrust law to ensure sustained contestability and fairness in evolving digital markets.

## 1. Introduction

Starting March 7, 2024, the European Union will regulate numerous platform services offered by Big Tech companies under the Digital Markets Act (DMA).[1] Gatekeepers, as designated by the Commission,[2] must comply with an extensive set of obligations. These rules aim to foster contestability and fairness in digital markets, thus enhancing both choice and innovation.

---

[1] Regulation 2022/1925 of the European Parliament and of the Council of 14 September 2022 on contestable and fair markets in the digital sector and amending Directives 2019/1937 and 2020/1828 (Digital Markets Act), (OJ L 265, 12.10.2022, p. 1) (DMA hereafter).

[2] See European Commission, Commission designates six gatekeepers under the Digital Markets Act, Press Release: 6 September 2023.

The DMA is often viewed through a retrospective lens, due to an undeniable continuity between the enforcement of competition laws against abuses of dominant positions and the obligations included in this new approach to ex-ante regulation. Consider, for instance the anti-steering provision of Article 5(4) of the DMA, which obliges gatekeepers to allow business users of their core platforms services to advertise to end users the existence of alternative subscription and billing mechanisms. Just days prior to the compliance deadline for the first six designated gatekeepers, the Commission issued an infringement decision to Apple for abusing its dominant position, specifically in relation to anti-steering practices in its App Store —practices now also prohibited by the DMA.[3]

What is less examined, or even dismissed out of hand,[4] is the DMA's ability to confront not only past but also current and future obstacles to the contestability and fairness of digital markets within the EU.[5] An interesting opportunity for early 'beta-testing' the DMA in this more future-oriented dimension is provided by a recent horizon scanning effort conducted by the UK competition authority.[6] Among other observations, the Competition and Markets Authority (CMA) identified a trend among digital platforms to expand their offerings by integrating additional services for their users, including social media platforms that incorporate in-app shopping and financial services.[7] This trend might also serve as a preliminary sign of the emergence of 'super-apps' in the UK similar to WeChat in China and KakaoTalk in South Korea.

A pertinent question arises regarding the impact of increased platform integration within the EU, under the DMA framework, especially since several potential super-app candidates—such as WhatsApp, Instagram, and TikTok— are already regulated as core platform services.[8] Ofcom, the UK telecom regulator, points specifically to mandated interoperability of messaging services, one of DMA obligations, as possibly mitigating competitive issues raised by super-apps.[9] In a similar vein, this paper investigates whether the DMA provides adequate safeguards precluding, or at least mitigating, the emergence of new risks to the contestability and fairness of digital markets from additional platform integration by certain gatekeepers. Informed by the impressive growth trajectories of WeChat and KakaoTalk, two messaging services, focus is placed on Meta's WhatsApp as a designated gatekeeper poised to potentially expand the platform's service offerings. This possible future scenario prompts a critical examination of whether the DMA is sufficiently forward-looking to safeguard fairness and contestability in the face of such potential developments.

---

[3] European Commission, Commission fines Apple over €1.8 billion over abusive App store rules for music streaming providers, Press Release, 4 March 2024.
[4] Cfr. Pablo Ibáñez Colomo, 'Future-Proof Regulation against the Test of Time: The Evolution of European Telecommunications Regulation' (2022) 42 Oxford Journal of Legal Studies 2022, Vol. 42, 1170, 1192.
[5] The topic has clearly already emerged with regard to recent AI advancements, see 'Future proof' DMA will help handle AI surge, EU's Guersent says, mlex, 5 February 2024.
[6] CMA, Trends in Digital Markets: A CMA Horizon Scanning Report, 14 December 2023.
[7] Ibid, at 17 ff.

## 2. WhatsApp as gatekeeping messaging service

Currently in the EU, there are two messaging services recognized under the DMA as core platform services provided by a gatekeeper, namely Meta's WhatsApp and Messenger.[10] Following a market investigation, the Commission has in fact decided that Apple should not designated as gatekeeper for its messaging service iMessage, despite being a core platform service and meeting the quantitative thresholds.[11]

According to the Commission, Meta fulfils the three cumulative requirements to be designated as a gatekeeper, namely: (a) it has a significant impact on the internal market; (b) it provides at least one core platform services (CPS) which is an important gateway for business users to reach end users; (c) it enjoys an entrenched and durable position, in its operations. Having in fact verified that the requisite quantitative thresholds are satisfied, the gatekeeper presumption laid down in Article 3(2) of the DMA is activated. Among the six Meta's CPSs identified by the Commission in its designation decision, there are two online social networking services (Facebook and Instagram), one online advertising service (Meta Ads), one online intermediation service (Marketplace), and two number-independent interpersonal communication services (NIICS), namely WhatsApp and Messenger.[12]

---

[8] Moreover, X (formerly known as Twitter) has disclosed the intention to become 'the everything app', see Elon Musk's 'everything app' plan for X, in his own words, The Verge, 31 October 2023. X recently notified to the Commission that it potentially meets the DMA thresholds, see Directorate-General for Competition, Directorate-General for Communications Networks, Content and Technology, Booking, ByteDance and X notify their potential gatekeeper status to the Commission under the Digital Markets Act, 1 March 2024.

[9] Ofcom, What super-apps could mean for the communications sector, 7 February 2024.

[10] Commission Decision of 5.9.2023 designating Meta as a gatekeeper pursuant to Article 3 of Regulation (EU) 2022/1925, C(2023) 6105 final (Meta Decision hereafter).

[11] See EC Press Release, 13 February 2024 (the non-confidential version of the Decision has not yet been published).

As explained by the Commission in its Meta designation decision, WhatsApp "allows end users to send text messages and to make voice and video calls to other individual WhatsApp end users or within WhatsApp groups, as well as to share pictures, videos, and other content." Moreover, the Commission has identified various business tools and interfaces provided by Meta in the context of the WhatsApp platform, such as WhatsApp Business App and WhatsApp for Business Application Programming Interface (API).[13] These tools, which are "designed to enhance the usage of WhatsApp", should be considered to form a part of the NIICS CPS WhatsApp.[14]

However, the services delivered by DMA gatekeepers are anticipated to evolve in response to technological advancements, business dynamics, and changing consumer demand. It is this very prospect that we try to address here, despite the intrinsic challenges associated with predicting the scope and extent of these services' evolution. Following the CMA's recognition of a possible trend towards platform integration and the emergence of super-apps, as discussed in the Introduction, Ofcom has also subsequently turned its attention specifically towards messaging services.[15] In Europe, the potential evolution of WhatsApp, by far the most popular messaging application,[16] through the integration of additional functions and services, could be foreshadowed by similar developments occurring in Brazil and India, where the app already integrates functionalities from marketing to in-app payments.[17] While it is conceivable that Meta would try to use WhatsApp as a platform expanding the services they offer their users along the lines of what happened in those countries, the age of AI offers new possibilities of vertical integration, as again spotted by the CMA. The UK authority mentions ChatGPT as an example of a chatbot powered by Foundation Models (FMs), which could become a new gateway for products and services by supporting and integrating plugins within the FM-based service.[18] It doesn't take a significant stretch of the imagination to foresee the integration of a similar FM-based service into the WhatsApp platform. In fact, this has already been done. Meta AI, a conversational chatbot, has been integrated into WhatsApp and other applications owned by Meta.[19]

---

[12] NIICS are defined by cross referencing to Article 2, point (7), of Directive (EU) 2018/1972 of the European Parliament and of the Council of 11 December 2018 establishing the European Electronic Communications Code (OJ L 321, 17.12.2018, p. 36).
[13] Cfr. Meta Decision, paragraph 130.
[14] Ibid, paragraph 131.
[15] Ofcom (n 9).

Given WhatsApp's significant and growing popularity also among business users, it is conceivable that AI and/or other technologies (like the metaverse) could be used by Meta in ways that extend beyond merely optimizing communication between end users and between end users and business users. In the scenario we very tentatively sketching, WhatsApp would be broadening its range of features through the introduction of new functionalities. Irrespective of the technological framework facilitating this integration, it is likely that these digital services and functionalities would be designed by Meta to be used in conjunction with, and "designed to enhance the usage" of WhatsApp.[20]

The UK telecom regulator highlights in particular that the current situation seems rather favourable to the users of WhatsApp and other messaging services, despite some competition risks.[21] However, depending on the concrete configuration of WhatsApp's platform integration, Ofcom admits that there could "new competitive concerns".[22] Given the logic underpinning the DMA's regulatory structure, which aims to comprehensively address and dismantle the strategies enabling digital platforms to secure an entrenched position in different business areas, thereby transforming digital markets and boosting innovation, the question naturally arises: is the DMA equipped to address new competitive concerns within the European Union, should they indeed materialize, thus preserving contestability and fairness going forward?

---

[16] See Bundeskartellamt, Sektoruntersuchung: Messenger- und Video-Dienste, Abschlussbericht, 15.05.2023, p.28.
[17] Ibid.
[18] CMA (n 6). ChatGPT Plugins beta, however, has already been discontinued and effectively replaced by the GPT store, see OpenAI, Winding down the ChatGPT plugins beta, https://help.openai.com/en/articles/8988022-winding-down-the-chatgpt-plugins-beta (last accessed: 8 March 2024).
[19] Cfr. Meta, Introducing New AI Experiences Across Our Family of Apps and Devices, Press Release, 27 September 2023.
[20] Cfr. Meta Decision, paragraph 131.

While it appears quite probable that some form of platform integration, potentially through the blending of AI or virtual worlds technologies in WhatsApp, may occur, determining the specific strategy Meta will actually implement is currently challenging. For this reason, in the following section, we will confine ourselves to highlighting those obligations applicable to WhatsApp that could, in fact, at least mitigate new contestability and fairness issues that might arise following platform integration. This, of course, in turn presupposes that the DMA is sufficiently and swiftly enforced.

## 3. Can the DMA obligations deliver?

The first thing to note is that platform integration by gatekeeping core platform services covered by the DMA is not prohibited per se. Other jurisdictions could be making different choices, such as Indonesia, which has recently banned commerce on social media platforms.[23] Another preliminary point to mention is that if the platform integration is achieved through a concentration (for example, by acquiring an FM developer), it could be subject to EU or national merger control. Moreover, Article 14 of the DMA requires the gatekeeper to inform the Commission of the concentration, irrespective of whether this is notifiable to the Commission under the EU Merger Regulation or to a competent national competition authority under national merger rules. In the following discussion, we will pinpoint the DMA obligations that could seem most relevant to our 'WhatsApp as would-be super-app scenario'.

---

[21] Ofcom, Personal online communication services: A discussion document, 25 October 2023.
[22] Ofcom (n 9).
[23] New York Times, TikTok Forced to Close Shopping Feature in Its Second-Largest Market, 5 October 2023.

The CMA explains that "the market for integrated platforms or super-apps has the potential to become concentrated within a single or very small number of platforms."[24] This is obviously a very forward-looking perspective, where super-apps have practically absorbed other so far standalone platform services. Instead of juggling multiple apps, the end user would prefer to access goods and services, or a specific selection thereof, through one or more super-apps. The emergence of FM-fuelled chatbots could lead to such a reorganization in the consumption of digital services. In other words, the super-app would be the new operating system. The circumstances that could lead to this new equilibrium are those we have come to know in the last few decades of increased familiarity with digital platforms: the ability to leverage the existing user base, direct and indirect network effects, data network effects, consumer biases, etc. Further aggravating elements would be the presence of switching costs, as well as the absence of multi-homing or horizontal interoperability.

In this context, super-apps could then resort to practices that raise the barriers to entry and expansion, further dissuade users from multi-homing, give favourable treatment to their own services, limit access to their functionalities, selectively disable or degrade some of those functionalities, make these functionalities available to business users only on the condition that they refrain from developing competing offerings, prevent business users from using competing platforms, etc.

Allegedly, the realization of this new equilibrium, on one hand, does not take into account the possibility of timing and effective interventions by antitrust authorities aimed at curbing anti-competitive practices executed by would-be super-apps, and on the other, assumes the absence of a specific ex-ante regulation such as the DMA.[25] While antitrust enforcement can be difficult to predict, we tentatively explore in the following what the impact of the new EU regulation on digital platforms might be on at least some of the very plausible, centrifugal dynamics and practices outlined above.

---

[24] CMA (n 6), p. 19.
[25] The latest available version of the UK Digital Markets, Competition and Consumers Bill can be found at https://bills.parliament.uk/publications/54208/documents/4421.

The initial point to observe regarding the potential evolution of WhatsApp into a super-app under the DMA umbrella is that this regulation mandates horizontal interoperability explicitly for this core platform service. This DMA obligation is also mentioned by Ofcom in its brief reflection on super-apps, as well as in its recent discussion document on personal communication services. As noted by the UK authority, "[a] lack of interoperability between competing services offered by different providers can give an advantage to larger platforms due to network effects" and "the case for intervention will tend to be stronger when a lack of interoperability is preventing users from choosing services based on price and quality considerations".[26] A similar observation is found in the DMA, which states that "the lack of interoperability allows gatekeepers that provide number-independent interpersonal communications services to benefit from strong network effects, which contributes to the weakening of contestability".[27] In scenarios where network effects are especially strong, the decision of users to opt for certain providers might be predominantly influenced by the user base's size, rather than considerations of price, quality, and functionality. This dynamic can diminish the competitive incentive for incumbents to maintain high standards of quality, to offer competitive pricing, and to roll out innovations that truly enrich user choice. Of course, this extends beyond just messaging services. In its investigation of online platforms and digital advertising, the CMA found that Facebook's market dominance is largely attributed to the significant network effects arising from its extensive base of connected users and the minimal interoperability it permits with other social media platforms.[28]

Article 7 of the DMA specifically targets reducing the impact of these firm-specific network effects, making it easier for users to adopt smaller platforms and potentially lowering the hurdles for entry or expansion for these smaller providers. However, it's important to emphasize in this respect that the DMA does not prevent companies already gatekeeping other core platform services from benefiting from the entry and expansion opportunities into somebody else's walled garden created by this new regulation. Article 7 of the DMA permits other messaging services to achieve interoperability with WhatsApp. This arrangement allows end-users of alternative messaging platforms, should they opt for interoperability, to send and receive messages to and from users of WhatsApp who have similarly opted in. In the first year, the mandate encompasses text messaging between individual users, along with the capability to share images, voice messages, videos, and other attached files amongst individual users. In subsequent years, the requirements will broaden to include group functionality and calling features.

---

[26] Ofcom (21), p. 5.
[27] DMA, Recital 64.
[28] CMA, Online platforms and digital advertising market study final report, July 2020, paragraph 91. However, the European Parliament's proposal to extend mandated interoperability, currently foreseen by the DMA for messaging services, to social networks did not receive support from the co-legislator.

While Article 7 of the DMA opens up the possibility for horizontal interoperability with WhatsApp, and even with the best possible implementation of this obligation, it will ultimately be the market that decides if there is indeed space for new messaging service providers, taking into account the myriad other market failures that could impede their emergence. And in any case, a horizontal interoperability obligation like the one introduced by the DMA is far from being a cure-all, considering the probable existence of market failure problems that persist regardless of the level of horizontal interoperability.[29] When these problems are serious, enhanced interoperability might have minimal impact on user behaviour without the implementation of complementary measures.[30] Some of these issues can be directly linked to the presence of barriers to multihoming and switching, which can at least partially be addressed through specific measures. For instance, the DMA contains measures aimed at tackling barriers to switching, such as Article 6(9), which requires Meta to provide end users, and third parties that they have authorised, with effective portability of data provided by the end user or generated through their activity on WhatsApp. Moreover, Article 6(10) requires Meta to provide business users - on request - with access to and use of data that is provided for, or generated in the context of the use of, WhatsApp by those business users and the end users that are engaging with the products or services provided by business users on WhatsApp. This would apply also to the extended functionalities that Meta would wish to integrate into this DMA core platform service.

---

[29] Ofcom (21), p.17.
[30] ibid, p.19.

Barriers to entry can be raised by the would-be super-app WhatsApp also in different ways, such as making using, accessing, signing-up or registering for Meta's designated CPSs conditional on subscribing to, or registering with, another of its designated CPSs. This is now prohibited by Article 5(8) of the DMA, as it would provide "a means of capturing and locking-in new business users and end users for their core platform services", as made clear by Recital 44.

The AI-driven super-app scenario previously outlined, which incorporates features like an FM-powered chatbot and the creation of some sort of 'store' for third party developers might also give rise to various vertical interoperability challenges. As AI-powered services increasingly become "must-have" services for users, the value of compatibility with other services will grow in importance from the users' perspective. WhatsApp could then tactically utilize these interoperability obstacles and challenges in ways reminiscent of strategies employed in app marketplaces by providers of operating systems. To counteract these pervasive problems with regard to designated software application stores, online search engines and online social networking services, the European legislator has specifically included vertical interoperability requirements in the DMA. Thus, Article 6(12) already requires Meta to apply fair, reasonable and non-discriminatory ("FRAND") for business users with respect to its online social networking services Facebook and Instagram, and to publish general conditions of access including an alternative dispute settlement mechanism. At the moment, however, this is not expressly foreseen for software application stores that are 'ancillary' to messaging services, as they might develop in the AI age. It will be necessary to assess, if needed, whether this becomes a regulatory gap that needs to be addressed in the future.

Other issues raised by Meta's dual role in the future in the 'AI store', in the 'metaverse' or elsewhere could also be tackled by specific obligations that already apply to Meta, such as Article 6(2) of the DMA. This provision requires Meta to ensure that it does not use non-public data provided or generated by business users in the context of their use of WhatsApp or of the services provided together with, or in support of, WhatsApp when that usage is "in competition with" such business users.

Various other DMA obligations come to mind that could apply to this future scenario, to be applied as they are or specified and adapted as provided by the Regulation. Nonetheless, given the uncertainty of the scenario, it's not possible to conclude that the DMA will be sufficient on its own. In any case, the enforcement of this new ex-ante framework will be complemented by traditional antitrust law.

## 4. Conclusion

The regulatory landscape for digital markets has significantly evolved with the rollout of the DMA, a pioneering global initiative in terms of its type, scope, and ambition. Although it builds on the foundation of antitrust law, the DMA sets out on a distinct path, emphasizing its unique role in shaping the future of digital market oversight. The relationship with traditional antitrust law is poised to evolve towards a more complementary nature. To assess whether the journey will be fruitful, a frequently debated aspect is the DMA's ability to be future proof, especially in light of the inevitable evolution of the core platform services it aims to regulate. In this contribution, we have sought to delve into this issue by envisioning a potential evolution of a gatekeeping messaging service. The brief analysis concludes that while the DMA can already address many of the contestability and fairness concerns arising from this type of platform integration, it must in turn be prepared to adapt to new changes as necessary and cultivate a fruitful complementarity with traditional antitrust law.